\documentclass[fleqn,usenatbib]{mnras}

\usepackage{newtxtext,newtxmath}

\usepackage[T1]{fontenc}

\DeclareRobustCommand{\VAN}[3]{#2}
\let\VANthebibliography\thebibliography
\def\thebibliography{\DeclareRobustCommand{\VAN}[3]{##3}\VANthebibliography}

\usepackage{graphicx}	
\usepackage{subcaption}
\usepackage{amsmath}
\usepackage{gensymb}

\title[Magnetic Field Tomography towards the L1688 Cloud]{Velocity Gradient and Stellar Polarization: Magnetic Field Tomography towards the L1688 Cloud}

\author[Schmaltz et al.]{
Tyler Schmaltz$^{1,2}$,
Yue Hu$^{1,2}$\thanks{E-mail: yue.hu@wisc.edu},
Alex Lazarian$^{1,2}$\thanks{E-mail:alazarian@facstaff.wisc.edu }
\\
$^{1}$Department of Physics, University of Wisconsin-Madison, Madison, WI, 53706, USA\\
$^{2}$Department of Astronomy, University of Wisconsin-Madison, Madison, WI, 53706, USA
}

\date{Accepted XXX. Received YYY; in original form ZZZ}

\pubyear{2023}

\begin{document}
\label{firstpage}
\pagerange{\pageref{firstpage}--\pageref{lastpage}}
\maketitle

\begin{abstract}
Magnetic fields are a defining yet enigmatic aspect of the interstellar medium (ISM), with their three-dimensional mapping posing a substantial challenge. In this study, we harness the innovative Velocity Gradient Technique (VGT), underpinned by magnetohydrodynamic (MHD) turbulence theories, to elucidate the magnetic field structure by applying it to the atomic neutral hydrogen (HI) emission line and the molecular tracer $^{12}$CO. We construct the tomography of the magnetic field in the low-mass star-forming region L1688, utilizing two approaches: (1) VGT-HI combined with the Galactic rotational curve, and (2) stellar polarization paired with precise star parallax measurements. Our analysis reveals that the magnetic field orientations deduced from stellar polarization undergo a distinct directional change in the vicinity of L1688, providing evidence that the misalignment between VGT-HI and stellar polarization stems from the influence of the molecular cloud's magnetic field on the polarization of starlight. When comparing VGT-$^{12}$CO to stellar polarization and Planck polarization data, we observe that VGT-$^{12}$CO effectively reconciles the misalignment noted with VGT-HI, showing statistical alignment with Planck polarization measurements. This indicates that VGT-$^{12}$CO could be integrated with VGT-HI, offering vital insights into the magnetic fields of molecular clouds, thereby enhancing the accuracy of our 3D magnetic field reconstructions.
\end{abstract}




\section{Introduction}
Magnetic fields serve as significant agents that drive the evolution and structure of the interstellar medium (ISM; \citealt{Crutcher12,2014A&A...561A..24B,BG15,refId0,2017ARA&A..55..111H,HYL20}). Their pivotal roles extend to various astrophysical processes, such as star formation \citep{MK04,Crutcher04,MO07,2012ApJ...761..156F,2012ApJ...757..154L,2014SSRv..181....1L,Hu_2022} and the propagation of cosmic rays \citep{1949PhRv...75.1169F,1966ApJ...146..480J,2002PhRvL..89B1102Y,2004ApJ...614..757Y,2010A&A...510A.101F,2013ApJ...779..140X,2020ApJ...894...63X,2021MNRAS.501.4184H,2021arXiv211115066H,2022ApJ...926...94M,2023JHEAp..40....1Z}. Particularly, when modeling the Galactic polarization foreground for detecting the polarization B-mode in the Cosmic Microwave Background (CMB), the Galactic magnetic fields are indispensable \citep{Pogosian_2014,Bracco_2019}.

Notwithstanding their significant role in the ISM, our comprehension of these magnetic fields is still incomplete. A primary challenge lies in the fact that the magnetic fields, when inferred from polarized dust emission, are projected onto the plane-of-the-sky (POS) and integrated along the line-of-sight (LOS; \citealt{10.1111/j.1365-2966.2007.11817.x,BG15,2016ApJ...831..159H,2020A&A...641A...3P,2023MNRAS.519.3736H,2023arXiv231017048H}). This results in the obfuscation of the magnetic fields' three-dimensional (3D) spatial distribution within the Galaxy. However, polarized starlight contains distance information along the LOS, primarily because the positions of stars can be pinpointed with precision, for example, by the GAIA satellite \citep{2023A&A...674A...1G}. This approach of combining stellar polarization and GAIA observation provides a promising way to reconstruct the Galactic magnetic field in 3D space with a sufficiently large sample of stars \citep{Fosalba_2002,2012ApJS..200...19C,2019ApJ...872...56P,González-Casanova_2019}. 

Recent advances in our understanding of MHD turbulence \citep{1995ApJ...438..763G, Lazarian_1999} and the nature of the 21 cm neutral hydrogen (H I) emission lines \citep{2000ApJ...537..720L,2023MNRAS.524.2994H} have paved another way for mapping magnetic fields in 3D. One such method is the velocity gradient technique (VGT; \citealt{2017ApJ...835...41G,Lazarian_2018,10.1093/mnras/sty1807}). This technique uses the velocity information provided by spectroscopic observations in conjunction with the Galactic rotation curve to recover the 3D spatial distribution of the HI gas within the Galaxy \citep{González-Casanova_2019,2023MNRAS.524.2379H}. The magnetic field orientation is then unveiled by MHD turbulence's anisotropy imprinted in HI's velocity channel maps. This means that the intensity structures perceived within narrow HI channels are predominantly governed by the turbulent velocity field and exhibit elongation along the magnetic fields due to the velocity caustic effect \citep{2000ApJ...537..720L,2023MNRAS.524.2994H}. The gradient within these intensity structures acts as an anisotropy detector and subsequently traces the magnetic field's orientation.

Efforts to reconstruct the Galactic magnetic field in 3D, employing the VGT, HI observation, and the Galactic rotational curve, have been at the forefront of several studies \citep{González-Casanova_2019,2023MNRAS.524.2379H}. For instance, \cite{2023MNRAS.524.2379H} delved into the potential of mapping both the magnetic field orientation and strength simultaneously. \cite{González-Casanova_2019} examined the synergy between VGT-HI and the aforementioned stellar polarization. Interestingly, \cite{González-Casanova_2019} observed that while there was general alignment between VGT-HI and stellar polarization, certain misalignments emerged, with the inferred magnetic fields from both approaches occasionally appearing perpendicular. One speculation posits that stellar polarization contains contributions from molecular clouds in which the magnetic fields might have changes in their direction, while VGT-HI is insensitive to the changes, resulting in this misalignment. However, this effect has not been investigated.

In this work, our aim is to explore two aspects: (1) the coherence of magnetic fields in the foreground/background and molecular clouds and (2) the alignment and misalignment between the stellar polarization, VGT-HI, as well as the magnetic field inferred from the molecular line $^{12}$CO with VGT. For this purpose, we target the low-mass star-forming region L1688 in the giant molecular cloud Ophiuchi A. L1688 is located at $\approx$ 139 pc away from the Sun \citep{2008AN....329...10M,2018ApJ...869L..33O}. Due to its active star formation, this region has been subject to multiple surveys at many wavelengths \citep{1989ApJ...340..823W,1991ApJ...379..683L,1992ApJ...395..516G,Bontemps_2001,2005A&A...438..661O,2015MNRAS.450.1094P}. \cite{2023MNRAS.524.4431H} has applied VGT to the $^{12}$CO and $^{13}$CO lines of L1688 and found a generally good agreement with the magnetic field inferred from Planck polarization at 353 GHz. In this study, we extend the analysis to include stellar polarization from multiple surveys \citep{2000AJ....119..923H, 2015ApJS..220...17K, 1976AJ.....81..958V, 1979AJ.....84..199W,10.1093/mnras/230.2.321} and reconstruct the magnetic field's 3D distribution in L1688's foreground and background via VGT-HI and the Galactic rotation curve. 

This paper is organized into six sections. \S~\ref{sec:data} details the observational data procured and employed in our analysis. The theoretical framework and procedural pipeline of the VGT are elucidated in \S~\ref{sec:method}. In \S~\ref{sec:results}, we present the results of magnetic fields derived from Planck polarization and as traced by VGT with $^{12}$CO and HI acting as tracers, followed by a comparative analysis with magnetic fields inferred from stellar polarization data. \S~\ref{sec:dis} delves into the discussion of the observed misalignment between VGT-HI and stellar polarization, as well as the integrative use of VGT-HI and VGT-$^{12}$CO to refine magnetic field tomography. The paper is summarized in \S~\ref{sec:con}.

\section{Observational Data}
\label{sec:data}
The region of our study is the L1688 sub-cloud of the giant molecular cloud Ophiuchi A. L1688 is chosen due to the prevalence of low-mass star formation and the large amount of stellar polarization and distance data available. L1688 has a very high star formation rate with 14–40 percent of molecular gas within the cloud in the process of star formation \citep{10.1111/j.1365-2966.2008.13750.x}. The cloud is relatively close, located at a distance 138.4 $\pm$ 2.6 pc from the Sun \citep{2008AN....329...10M,2018ApJ...869L..33O}. Characterization of the magnetic field and associated self-gravity were conducted in an earlier study \citep{2023MNRAS.524.4431H}. This study also found Alfv\'en Mach number to be less than one because magnetic field pressure is greater than the turbulent pressure in L1688.

\subsection{$^{12}$CO emission line}
In this work, we employ the $^{12}$CO (1-0) emission line provided by the COMPLETE survey \citep{2006AJ....131.2921R}. The data were acquired via an observation using the 14 m Five College Radio Astronomy Observatory (FCRAO) telescope. $^{12}$CO emission line has an effective velocity resolution of 0.07 km/s. The full width at half maximum (FWHM) The Half Power Beam Width (HPBW) of the $^{12}$CO observation is $\approx$ 46". However, the final data cube is convolved onto a regular 23" per pixel resolution to satisfy the Nyquist sampling. The RMS noise level per channel is $\approx$ 0.98 K for $^{12}$CO in unit of antenna temperature $T^{*}_A$. The radial velocity of the cloud’s bulk motion ranges from 0 to 7 km/s. This velocity range was chosen for our analysis in this work.

\subsection{HI Emission}
In this work, we obtained our HI data from the 100 m NRAO Green Bank Telescope (GBT) in West Virginia which observed Ophiuchus in 2011 \citep{2003ApJ...585..823L}\footnote{Data archive: \href{https://dataverse.harvard.edu/dataset.xhtml?persistentId=doi:10.7910/DVN/R7DYQP}{doi:10.7910/DVN/R7DYQP}}.
The HI data utilized from the GBT has a pixel size or effective resolution of $1'$, an angular resolution of $\sim9'$, a spectral resolution of 0.32 km s$^{-1}$, and a normal RMS value of 0.15 K per channel \citep{2006AJ....131.2921R}. 
In this work, we analyze the velocity range from -60 km/s to 30 km/s.

\subsection{Planck polarization}
In this study, we have employed the magnetic field orientations deduced from the Planck polarization data and conducted a comparative analysis with the VGT and stellar polarization. Specifically, we utilized the 353 GHz polarized dust signal data from the Planck 3rd Public Data Release (DR3) in 2018 of the High-Frequency Instrument \citep{2020A&A...641A...3P}.

The Planck observations determine the polarization angle, $\phi$, using the Stokes parameter maps for intensity $I$, and the polarization states $Q$ and $U$. The angle $\phi$ is mathematically defined by the following relation:
\begin{equation}
\phi = \frac{1}{2} \tan^{-1}(-U,Q),
\end{equation}
where the notation $-U$ is employed to align the angle with the IAU convention, differing from the HEALPix standard. To improve the signal-to-noise ratio, the Stokes parameter maps were convolved with a Gaussian kernel to smooth the data from the original angular resolution of 5$'$ to 10$'$. The magnetic field angle, denoted as $\phi_B$, is subsequently derived by $\phi_B = \phi + \pi/2$.

\subsection{Stellar polarization}
In this study, we have harnessed stellar polarization data from a variety of surveys to map the magnetic field \citep{2000AJ....119..923H, 2015ApJS..220...17K, 1976AJ.....81..958V, 1979AJ.....84..199W, 10.1093/mnras/230.2.321}. The polarization measurements for the stars included in our analysis were taken in the K band from each respective survey. The selection criteria for inclusion were a signal-to-noise ratio greater than 3, and a location within the bounds of the L1688 star-forming region.

Distance information for the stars is directly accessible from the survey conducted by \cite{2000AJ....119..923H} online. For the remaining surveys \citep{1976AJ.....81..958V, 1979AJ.....84..199W, 10.1093/mnras/230.2.321}, which only provide data in paper-table form, we have compiled the relevant information—including stars' coordinates, polarization angle, and distance—into Tab.~\ref{tab:1} for ease of reference.

\section{Methodology}
\label{sec:method}
\subsection{VGT Basics}
VGT is a new method of mapping magnetic fields in the ISM theoretically based on MHD turbulence theories \citep{1995ApJ...438..763G} and fast turbulent reconnection theory \citep{Lazarian_1999}. It has been applied in the study of magnetic fields in a multitude of interstellar clouds, including neutral clouds, ionized clouds, and molecular clouds. \citep{HLY20,2020MNRAS.496.2868L,Hu_2021b,2022arXiv220512084T,2022MNRAS.511..829H}.  

Magnetic fluctuations are anisotropic, which was originally proposed in \citep{1995ApJ...438..763G}(denoted as GS95 hereafter). GS95 also discussed the relation of the anisotropy with the "critical balance" condition, where the cascading time $(k_{\perp}v_l)^{-1}$ must be equal to the wave period $(k_{\parallel}v_A)^{-1}$. $k_{\perp}$ and $k_{\parallel}$ denote wavevectors measured is 
perpendicular and parallel to the magnetic field. $v_A$ is defined as Alfv\'en velocity and $v_l$, where $v_l \propto l^{1/3}$ is defined as turbulent velocity, accounting for Kolmogorov-type turbulence. From this, the turbulent eddies of the magnetic field can be shown to have anisotropic properties where they elongate along the magnetic field. The scaling presented in originally in \cite{1995ApJ...438..763G} (henceforth GS95) in the form of wavenumbers, must be rewritten to account for the anisotropy of magnetic field that must be evaluated locally, rather than in respect to the mean magnetic field direction. This can be understood from the theory of turbulent reconnection,\footnote{The time taken for the reconnection of eddies should be mentioned as equal to the eddy turnover time, which was predicted by fast turbulent reconnection \citep{Lazarian_1999}. 
Consequently, the magnetic field doesn't stop the evolution of these eddies, and almost all of the energy of the cascade is channeled into them \citep{2023arXiv230106709Z}. } and it corresponds to numerical testing \citep{Lazarian_1999, 2000ApJ...539..273C}
\begin{equation}
\label{eq.gs95}
\begin{aligned}
         \textit{$k_{\parallel}$}  \propto   \textit{$k_{\perp}^{2/3}$},
\end{aligned}
\end{equation}
In the global system of reference related to the mean magnetic field, largest scale eddies dominate the anisotropy making it scale-independent. From \citep{Lazarian_1999}, it was found that eddies obey the hydrodynamic Kolmogorov scaling because magnetic fields give little resistance to eddy motion that is perpendicular to the local direction of the magnetic field. This scaling law can be defined as $v_{l,\bot}\propto l_{\bot}^{1/3}$ where $v_{l,\bot}$ is the turbulence's component which is perpendicular to the local magnetic field at scale $l$. Additionally, we know because eddies are affected only by the local magnetic field, both $l_\parallel$ and $l_\perp$ should be established with regards to the magnetic field's local direction. The previously mentioned critical balance condition can be modified and it can be shown $l_{\perp}/v_l$) should be equal to the wave period ($l_{\parallel}/v_{\rm A}$). 

From this, the relationship between the parallel and perpendicular scales of the eddies in the local reference frame can be acquired \citealt{Lazarian_1999}):
\begin{equation}
\begin{aligned}
l_{\parallel}  =  L_{\rm inj}(\frac{l_{\bot}}{L_{\rm inj}})^{2/3}M_{\rm A}^{-4/3},~M_{\rm A} \le 1,
\end{aligned}
\end{equation}
where $L_{\rm inj}$ denotes the turbulent injection scale, $M_{\rm A}=v_{\rm inj}/v_A$ denotes the Alfv\'en Mach number, and $v_{\rm inj}$ denotes the turbulence's injection velocity. Because of the anisotropy $l_\bot\ll l_\parallel$, the amplitude of velocity fluctuation and its corresponding gradient can be found via:
\begin{equation}
        \begin{aligned}
           &v_{l,\bot}  =  v_{\rm inj}(\frac{l_{\bot}}{L_{\rm inj}})^{1/3}M_{\rm A}^{1/3},~M_{\rm A} \le 1,\\
          &\nabla v_l\propto \frac{v_{l,\bot}}{l_{\bot}}=\frac{v_{\rm inj}}{L_{\rm inj}}M_{\rm A}^{1/3}(\frac{l_{\bot}}{L_{\rm inj}})^{-2/3},~ M_{\rm A} \le 1,
       \end{aligned}
\end{equation}
Here, $v_{\rm inj}$ refers to injection velocity. The $\nabla v_l$ here points in the direction of maximum variation in the velocity's fluctuation amplitude which will be perpendicular to the local magnetic field. Thus, this provides a convenient way to map magnetic fields.

\subsection{VGT Pipeline}
In this work, we utilize the Velocity Channel Gradients (VChGs). VChGs work by using a thin velocity channel, which is mapped as Ch($x$,$y$). Fluctuations in these velocity channels in Postion-Position-Velocity (PPV) space were theorized in \cite{2000ApJ...537..720L}. It was shown, that the narrower the velocity channel width, the greater the contribution of the velocity fluctuations is \citep{2023MNRAS.524.2994H}. When the velocity channel width $\Delta v$ is smaller than the velocity dispersion $\sqrt{\delta (v^2)}$ of the turbulent eddies being studied, then the intensity fluctuation in the thin channel will be dominated by velocity fluctuations. Conversely, in a thick channel, the intensity fluctuation will be dominated by density fluctuation. The thick and thin channels can be defined by the following criteria:
\begin{equation}
    \begin{aligned}
    \Delta v < \sqrt{\delta(v^2)} \text{, thin channel}\\
    \Delta v \geq \sqrt{\delta(v^2)}\text{, thick channel}
    \end{aligned}
\end{equation}

The minimum value of velocity channel width $\Delta v$ is constrained by the spectral or velocity resolution of the observation, while it is possible to combine several channels to get larger $\Delta v$. $\sqrt{\delta(v^2)}$ is typically calculated from the dispersion of the velocity centroid map, i.e., the moment-one map.

We can calculate the gradient map mathematically from these equations: $\psi_{\rm g}$:
\begin{equation}
\label{eq.5}
\begin{aligned}
\nabla_x{\rm Ch}_i &= {\rm Ch}_i(x,y) - {\rm Ch}_i(x-1,y),\\
\nabla_y{\rm Ch}_i &= {\rm Ch}_i(x,y) - {\rm Ch}_i(x,y-1),\\
\psi_{g}^{i}  &= {\rm tan}^{-1}(\frac{\nabla_y{\rm Ch}_i(x,y)}{\nabla_x{\rm Ch}_i(x,y)}),
\end{aligned}
\end{equation}
In these equations, $\nabla_{x}$Ch$_{i}(x,y)$ and $\nabla_{y}$Ch$_{i}(x,y)$ denote the $x$ and $y$ components of the gradients in the thin channel map, respectively. The subscript of these equations, $i = 1,2,...,n_v$, indicates the $i^{\rm th}$ channel map where $n_v$ is the total number of velocity channels. These equations are utilized and applied to each pixel that has a spectral line emission with a signal-to-noise ratio $>$ 3.

Even with these equations, the gradient orientation applied to each pixel, can statistically derive turbulence's property. The gradient map $\psi_{\rm g}$ is further processed via the sub-block avveraging method introduced in \cite{2017ApJ...837L..24Y}. The gradient map is divided into sub-blocks and a Gaussian fitting is applied within each sub-block to find the most probable orientation of gradient. The peak value of the each Gaussian distribution is utilized as the mean gradient for each corresponding specific sub-block. The processed gradient map is thus denoted as $\psi^{i}_{gs}(x,y)$.

Similar to Planck polarization measurements, we formulate the Pseudo-Stokes-parameters $Q_g$ and $U_g$ from $\psi^{i}_{gs}(x,y)$ using the following equations:
\begin{equation}
\begin{aligned}
Q_g(x,y) &= \sum_{i=1}^{n_v} I_i(x,y){\rm cos}(2\psi^{i}_{gs}(x,y)),\\
U_g(x,y) &= \sum_{i=1}^{n_v} I_i(x,y){\rm cos}(2\psi^{i}_{gs}(x,y)),\\
\psi_g &= \frac{1}{2}{\rm tan}^{-1}(\frac{ U_g}{Q_g}),
\end{aligned}
\end{equation}
where the pseudo polarization angle is represented by $\psi_g$ and is perpendicular to the POS magnetic field. The magnetic field orientation can then be inferred from orthogonality to be $\psi_B = \psi_g + \pi/2$

\begin{figure}
	\includegraphics[width=1.0\linewidth]{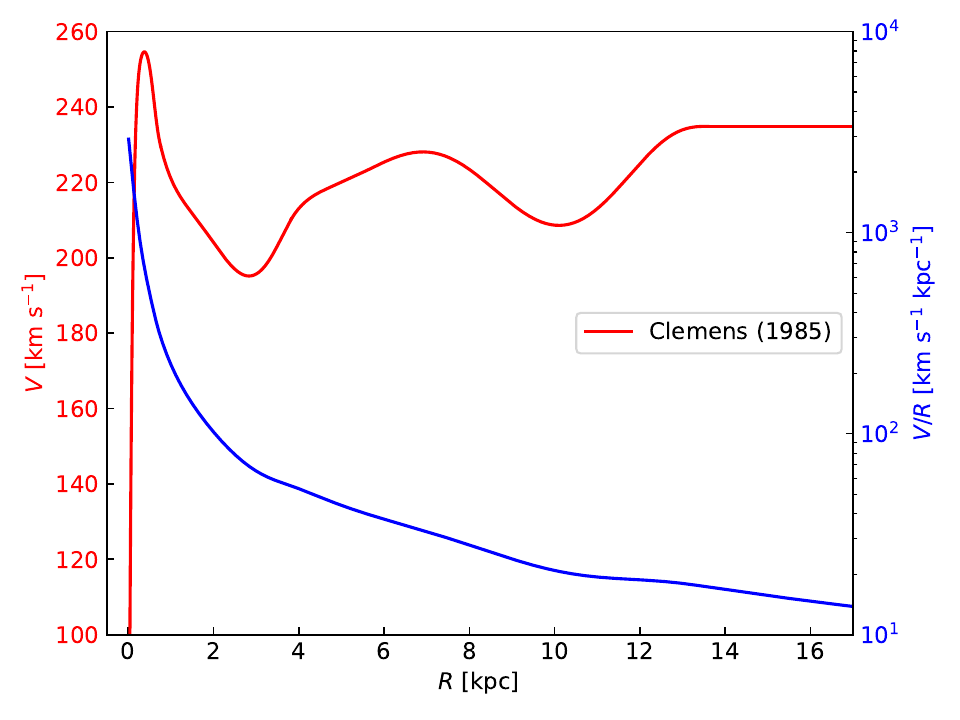}
    \caption{\textbf{Red:} Galactic rotation curve adopted from \citep{1985ApJ...295..422C}. \textbf{Blue:} The ratio $V/R$ calculated using $V$ from the the Galactic rotation curve. Reproduced from \citet{2023MNRAS.524.2379H}.}
    \label{fig:Velocity Rotation Curve}
\end{figure}

\subsection{Galactic rotational curve}
In order to acquire a spatial distribution of HI gas, an accurate Galactic rotational curve of the Milky Way Galaxy is required. In this work, we utilize the high-order polynomial curve obtained by \citep{1985ApJ...295..422C}. The equation of the rotational curve is represented in the following lines:

\begin{equation}
   \begin{aligned} 
V(R)&=\sum^{6}_{i=0}A_{i}R^{i},\quad R<0.09R_0\\
&= \sum^{5}_{i=0}B_{i}R^{i},\quad 0.09R_0<R<0.45R_0\\
&= \sum^{7}_{i=0}C_{i}R^{i},\quad 0.45R_0<R<1.6R_0\\
&=D_0,\quad 1.6R_0<R,
    \end{aligned}
\end{equation}
where $R$ denotes the distance from a point of interest in the Galaxy to the Galactic center, $R_0=8.5$ kpc and denotes the distance from the Galactic center to the Sun in our solar system, and $V$ denotes the circular velocity of the point of interest. $A,B,C,$ and $D$ are fitted coefficients of the curve using the assumption that our Sun's circular velocity is $V_0=220$ km/s. The specific values of these coefficients can be found in Table 3 in \citep{1985ApJ...295..422C}. Rather than listing the numerous coefficients, we reproduce the rotational curve as shown in Fig.~\ref{fig:Velocity Rotation Curve}. After 1.6$R_0$, we assume that the curve will flatten.

Using the curve, the spatial distribution of the HI gas at the point of interest can easily be found using the relative velocity $V_r$ and the angular velocity $\omega=V/R$ of the Galactic rotation. Relative velocity can be expressed as \citep{McClure-Griffiths_2007}:
\begin{equation}
\begin{aligned}
    V_R=R_0(\frac{V}{R} - \frac{V_0}{R_0})\sin{l}\cos{b},
\end{aligned}
\end{equation}
where $l$ denotes Galactic longitude and $b$ denotes Galactic latitude in the Galactic coordinate system. For $l$ and $b$, we use the central coordinates of the L1688 cloud. We determine $V_r$ via the central velocity of each HI channel. 
\begin{figure}
    \includegraphics[width=1\linewidth]{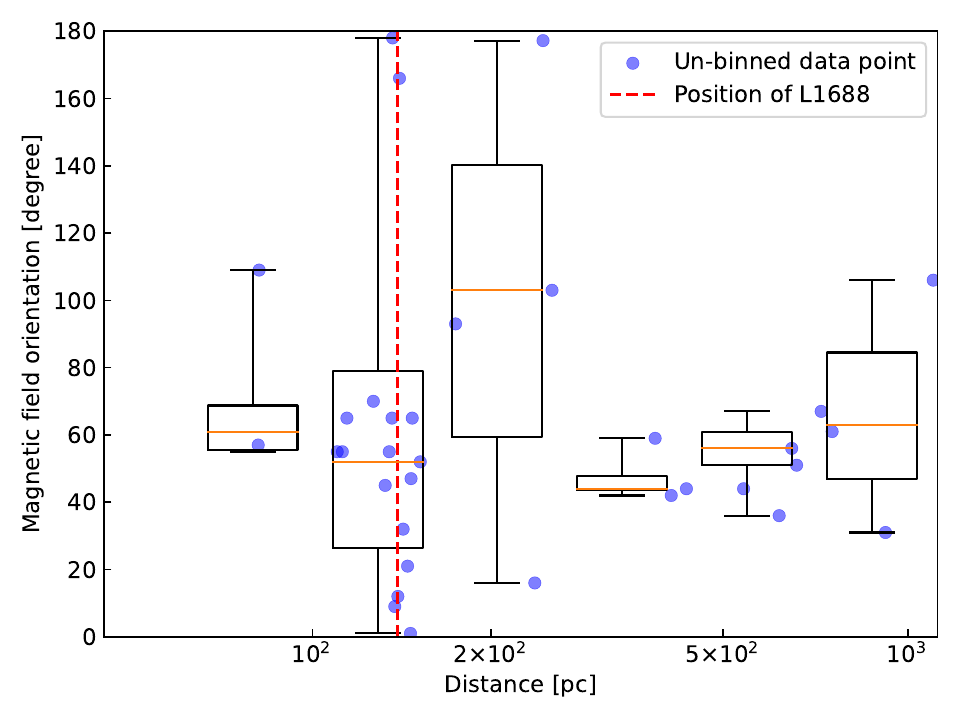}
    \caption{The figure presents the magnetic field orientation (east of the north) mapped with stellar polarization versus the corresponding stars' distances along the LOS. The upper and lower black lines represent the deviation’s maximum and minimum, respectively. The box gives ranges of the first (lower) and third quartiles (upper) and the orange line represents the median value.}
    \label{fig:Polarization Angle vs. Distance}
\end{figure}
\section{Results}
\label{sec:results}
\subsection{Change of magnetic fields orientation mapped with stellar polarization}
Fig.~\ref{fig:Polarization Angle vs. Distance} delineates the orientation of the magnetic field as a function of stellar distances. Predominantly, the magnetic field angles span between [30$^\circ$, 110$^\circ$], with the median orientation settling at $60^\circ$ east of the north. Notably, a significant shift is observed at the location corresponding to L1688, approximately 139 pc away and up to 200 pc, where the angle exhibits a bifurcation—either escalating to nearly $180^\circ$ or plummeting to $0^\circ$. This bifurcation is indicative of a potential difference in the magnetic field orientation between the foreground/background and that within the L1688 cloud itself.

To ascertain the relative contributions of the foreground/background and the molecular cloud, we analyze their hydrogen column densities, assuming dust grains are uniformly mixed with hydrogen. Instead of using the typical SED fitting method, the column density of hydrogen for the foreground/background, denoted as $N_{\rm H}^{\rm f}$, is calculated using the X-factor method from \citep{2019ApJ...872...56P}:
\begin{equation}
\label{eq.10}
\begin{aligned}
N_{\rm H}^{\rm f}&=N_{\rm H I}+2N_{\rm H_2}\approx N_{\rm H I}\\
&=\int 1.823\times10^{18}T_{\rm MB}^{\rm HI} dv~\text{cm}^{-2}\text{/(K km s}^{-1}\text{)},
\end{aligned}
\end{equation}
where $N_{\rm HI}$ represents the column density of atomic hydrogen (H I), $N_{\rm H_2}$ denotes the column density of molecular hydrogen (${\rm H_2}$), and $T_{\rm MB}^{\rm HI}$ signifies the brightness temperature of H I emission.

For the molecular cloud L1688, the hydrogen column density is derived similarly using the CO-${\rm H_2}$ conversion factor $X_{\rm CO} = 2\times10^{20}$, commonly applied in giant molecular clouds \citep{2011MNRAS.418..664N}:
\begin{equation}
\begin{aligned}
N_{\rm H}^{\rm mc}&=N_{\rm H I}+2N_{\rm H_2}\approx 2N_{\rm H_2}\\
&= 2X_{\rm CO}(\int T_{\rm MB}^{\rm CO} dv)~\text{cm}^{-2}\text{ / (K km s}^{-1}\text{)},
\end{aligned}
\end{equation}
where $T_{\rm MB}^{\rm CO}$ is the brightness temperature of CO emission. Differing from Eq.~\ref{eq.10}, the contribution of H I is considered negligible here as ${\rm H_2}$ emission predominates.

Our findings (see Appendix \ref{app:D}) indicate that the average values of $N_{\rm H}^{\rm f}$ and $N_{\rm H}^{\rm mc}$ are comparably significant, with $N_{\rm H}^{\rm f}\approx2\times10^{21}{\rm cm}^{-2}$ and $N_{\rm H}^{\rm mc}\approx7\times10^{21}{\rm cm}^{-2}$. This suggests that both the foreground/background and the molecular cloud contribute notably to stellar polarization. At a distance of approximately 139~pc, the molecular cloud's contribution predominates, significantly affecting the direction of stellar polarization to reflect the magnetic fields within the cloud. Conversely, at larger distances, the background's contribution gradually increases, leading to stellar polarization being influenced by both the foreground/background and the molecular cloud.

\subsection{VGT-HI: magnetic fields in the foreground and background}
\begin{figure}
    \centering
    \includegraphics[width=1.0\columnwidth]{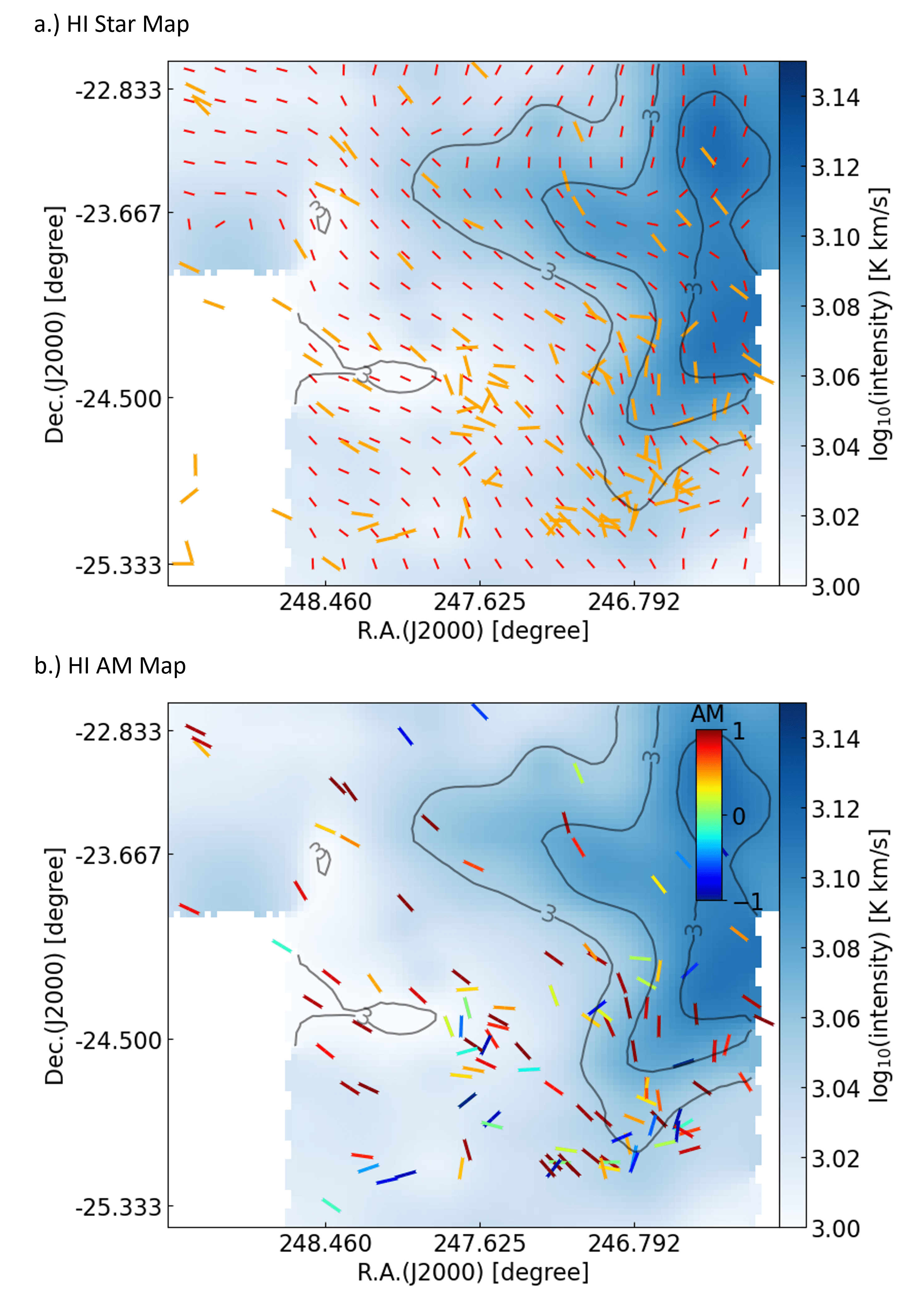}
    \caption{The top figure presents a mapping of star positions and their polarizations over the magnetic field orientation inferred by VGT, using total integrated HI as a tracer. The bottom figure presents the AM measurement comparing the magnetic field mapped by VGT versus the magnetic field inferred by stellar polarization.  The contours start from $\log_{10}(3.0)$ K km/s.
    }
    \label{fig:Hi Integrated maps}
\end{figure}

\begin{figure}
    \centering
    \includegraphics[width=1.0\columnwidth]{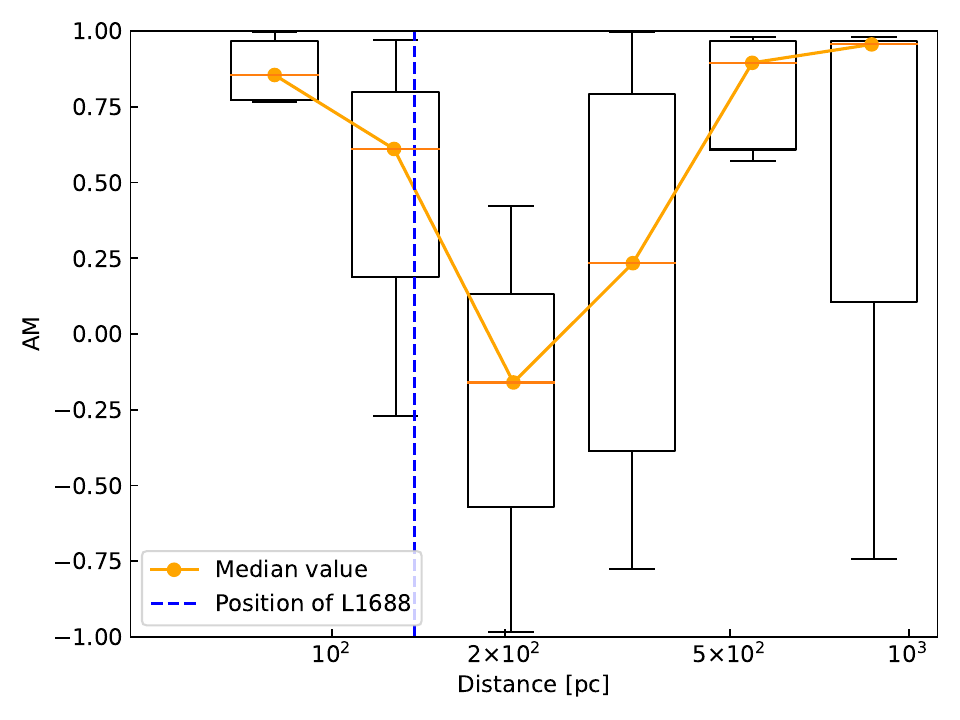}
    \caption{This figure presents the mean AM values of the stellar polarization at certain distances compared to VGT-HI. The upper and lower black lines represent the deviation’s maximum and minimum, respectively. The box gives ranges of the first (lower) and third quartiles (upper) and the orange line represents the median value.}
    \label{fig:HI_AM_distance}
\end{figure}

\begin{figure*}
    \centerline
    {\includegraphics[width=1.\linewidth]{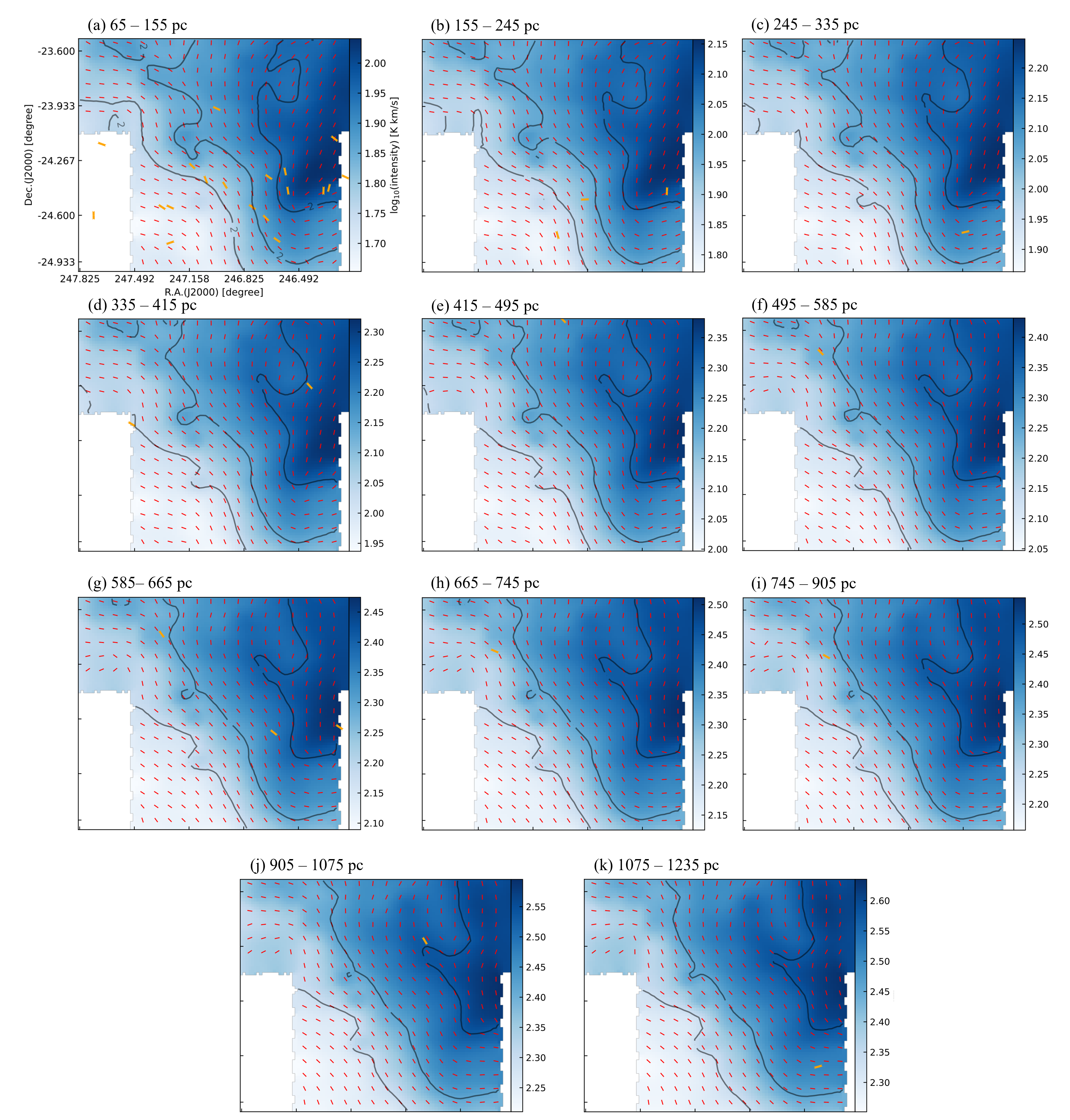}}
    \caption{The top figure presents a mapping of star positions and their corresponding polarization measurement over the magnetic field orientation inferred by VGT, using total integrated HI as a tracer. The bottom figure presents the AM measurement comparing magnetic field mapped by VGT versus the magnetic field inferred by stellar polarization.}
    \label{fig:HI Velocity Channels} 
\end{figure*}

\begin{figure}
    \centering
    \begin{subfigure}[b]{1.0\linewidth}
        \includegraphics[width=1.0\linewidth]{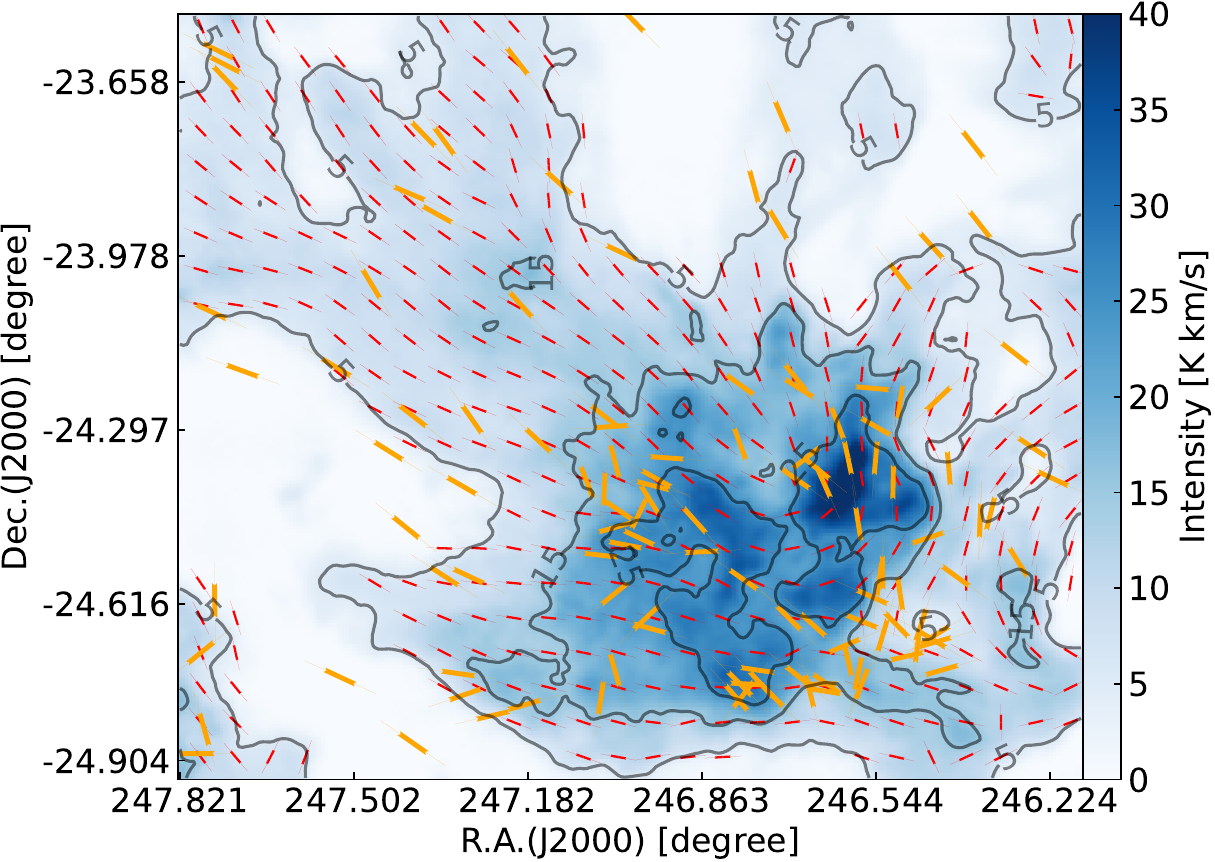}
        \caption{VGT-$^{12}$CO}
        \label{fig:sub1}
    \end{subfigure}
    \hfill
    \begin{subfigure}[b]{1.0\linewidth}
        \includegraphics[width=1.0\linewidth]{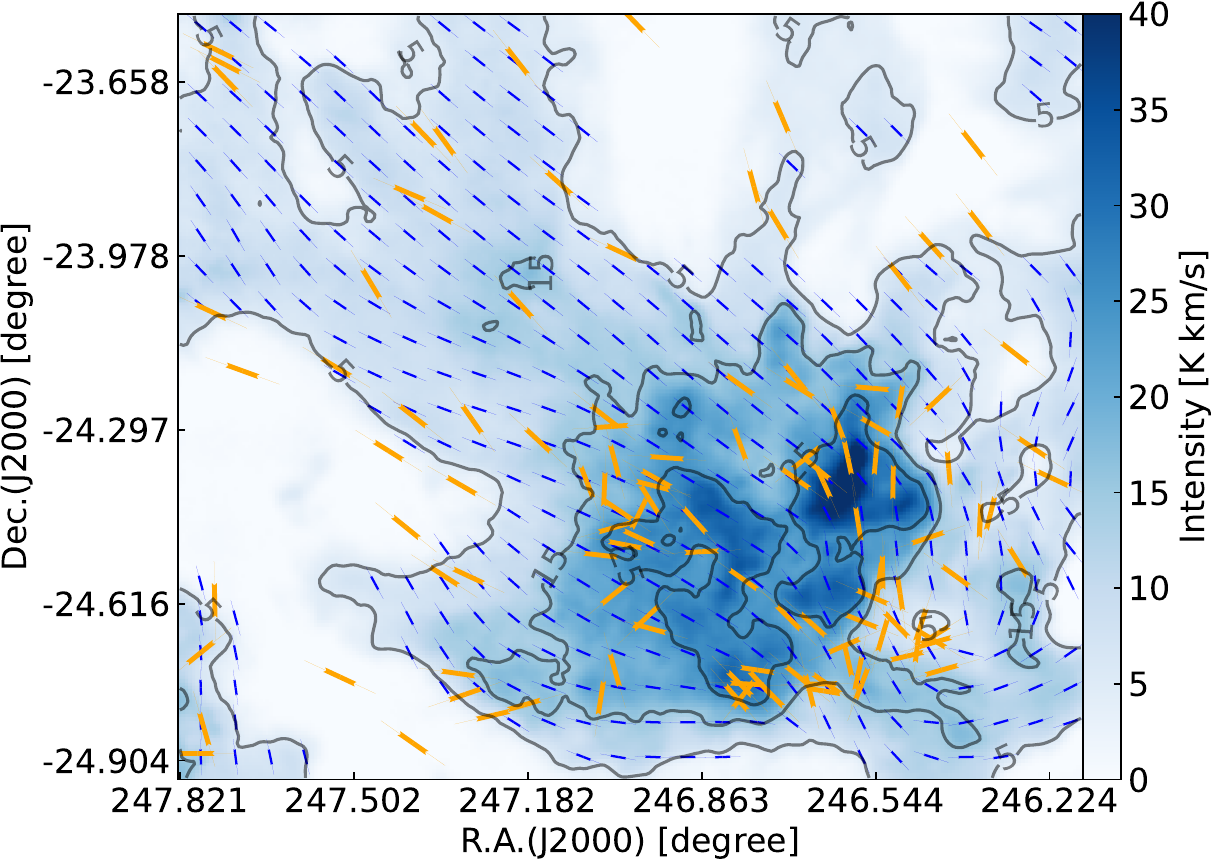}
        \caption{Planck polarization}
        \label{fig:sub2}
    \end{subfigure}
    \caption{Mapping of star positions and their polarization angles on top of the magnetic field orientation inferred from the VGT using $^{12}$CO (top) and the magnetic field inferred by Planck polarization (bottom). Contours for $^{12}$CO and Planck maps are both 
    5, 15, and 25 K km/s.}
\label{fig:VGT Star Maps}
\end{figure}

The VGT has proven to be effectively applicable to atomic HI emissions. When compared to molecular emissions, the VGT applied to HI (VGT-HI) has the distinctive advantage of mapping the magnetic fields that exist in the cloud's foreground and background \citep{González-Casanova_2019,2020MNRAS.496.2868L,2023MNRAS.524.2379H}. In Fig.~\ref{fig:Hi Integrated maps}, we compare the integrated VGT-HI map with stellar polarization data \footnote{We plotted all available stellar polarization data from multiple surveys \citep{2000AJ....119..923H, 2015ApJS..220...17K, 1976AJ.....81..958V, 1979AJ.....84..199W, 10.1093/mnras/230.2.321}, including data for stars without available parallax measurements. Figs.~\ref{fig:HI_AM_distance}, \ref{fig:HI Velocity Channels}, and \ref{fig:AM_distance} exclusively use data from stars with known distance information.}. This comparison extends beyond the stars for which distance data are available (as depicted in Fig.~\ref{fig:Polarization Angle vs. Distance}) to include those without known distances. It is observed that the global magnetic field patterns deduced from VGT-HI broadly align with those inferred from stellar polarization. Nevertheless, notable differences are present in the southern reaches of the map, especially the central dense clump of L1688 (see Fig.~\ref{fig:VGT Star Maps}). Given the magnetic field variations in L1688 identified in Fig.~\ref{fig:Polarization Angle vs. Distance}, it is plausible that the stellar polarization is significantly influenced by the molecular cloud, accounting for the observed misalignments.

Furthermore, to quantify our comparison between the magnetic field inferred by stellar polarization and VGT-HI, we use the Alignment Measure (AM; \citealt{2017ApJ...835...41G}), expressed as:
\begin{equation}
\begin{aligned}
{\rm AM} = \cos(2\theta_{\rm r}),
\end{aligned}
\end{equation}
here, $\theta_{\rm r} = \vert\phi_{\rm B}-\psi_{\rm B}\vert$. An AM value of 1 implies parallel alignment of $\phi_{\rm B}$ and $\psi_{\rm B}$, while -1 indicates perpendicular alignment. The AM map presented in Fig.~\ref{fig:Hi Integrated maps} demonstrates a concentration of positive AM values in the northeastern, lower-intensity tail of the cloud, whereas regions of negative AM predominantly coincide with the central dense clump.

Fig.\ref{fig:HI_AM_distance} illustrates the variation in AM as a function of distance along the LOS to the stars. It is crucial to note that the VGT-HI maps from which these measurements are derived represent an integration along the LOS. We observe that the AM is predominantly positive at smaller distances (approximately $<200$~pc), which is indicative of the foreground magnetic field alignment. Beyond 200 pc, where the influence of the L1688 cloud becomes more pronounced, there is a tendency for the AM to increase again at larger distances ($>200$~pc). An evident decrease in AM to negative values occurs around 
$\sim200$~pc. The data is binned uniformly in logarithmic space, and the bins are of equal size in non-logarithmic space, resulting in a physical bin interval of $\sim80$~pc between the second and third bins from the left, as shown in Fig.~\ref{fig:Polarization Angle vs. Distance}. The bin at $\sim200$~pc, thus, should contain also the contribution from the cloud. 

As we have previously discussed, this decline in AM corresponds to distances along the LOS that are near or slightly beyond the cloud. In Fig.~\ref{fig:Polarization Angle vs. Distance}, we note that the polarization angles for stars in the vicinity of L1688 range widely from 0 to 180 degrees. In contrast, the polarization angles for the remainder of the stars are generally confined between approximately 30 to 110 degrees with a median value of 60 degrees. Such a variation within the magnetic field orientations associated with a molecular cloud, however, would not be captured by VGT-HI due to the lack of HI in the cloud, where H$_2$ and other molecules are dominant.

\subsection{VGT-HI: magnetic fields in 3D}
\label{sec:foreground and background}
To facilitate a more direct comparison between VGT-HI mapping and stellar polarization, we employed the Galactic rotation curve method detailed in \S~\ref{sec:method} to ascertain the spatial positioning of each HI channel within the Galaxy. Fig.~\ref{fig:HI_spectrum} displays the original HI spectrum alongside its corresponding distance from the Galactic center.

In Fig.~\ref{fig:HI Velocity Channels}, stars are superimposed onto 11 distinct velocity channels, each aligned with their respective distances from the Sun. Across these channels, we observe subtle variations in the magnetic field, reflecting the nuanced 3D magnetic field mapping that polarization studies alone had previously found challenging. A closer examination reveals that, compared to the integrated HI map, there is a more pronounced agreement between the magnetic field gradient orientations and the stars within each channel. Most notably, there is a consistent positive correlation between the VGT-HI-derived magnetic field orientations and those of the stars at comparable distances. Exceptions are present in a few channels—for instance, sub-figures a (65 - 155 pc), b (155 - 335 pc), d (335 - 585 pc), and g (585 - 665 pc)— have some stars which exhibit discrepancies, typically within high-intensity regions. Nevertheless, these instances of disagreement may be attributed to the limited number of stars in those samples. When examining larger samples, the statistical agreement improves, reinforcing the effectiveness of VGT-HI in tracing magnetic field orientations across different Galactic scales.

\begin{figure}
    \centering
    \begin{subfigure}[b]{1.0\linewidth}
        \includegraphics[width=1.0\linewidth]{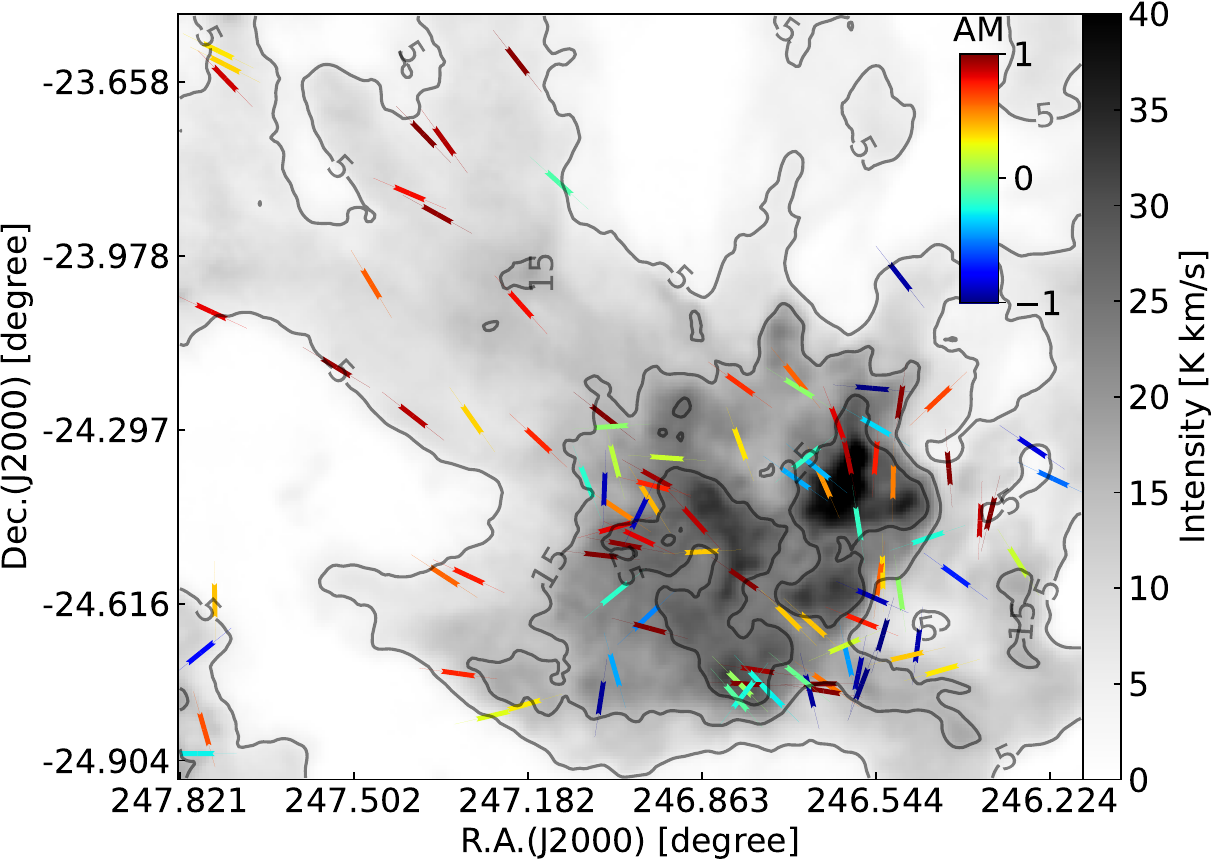}
        \caption{VGT-$^{12}$CO}
        \label{fig:sub1}
    \end{subfigure}
    \hfill
    \begin{subfigure}[b]{1.0\linewidth}
        \includegraphics[width=1.0\linewidth]{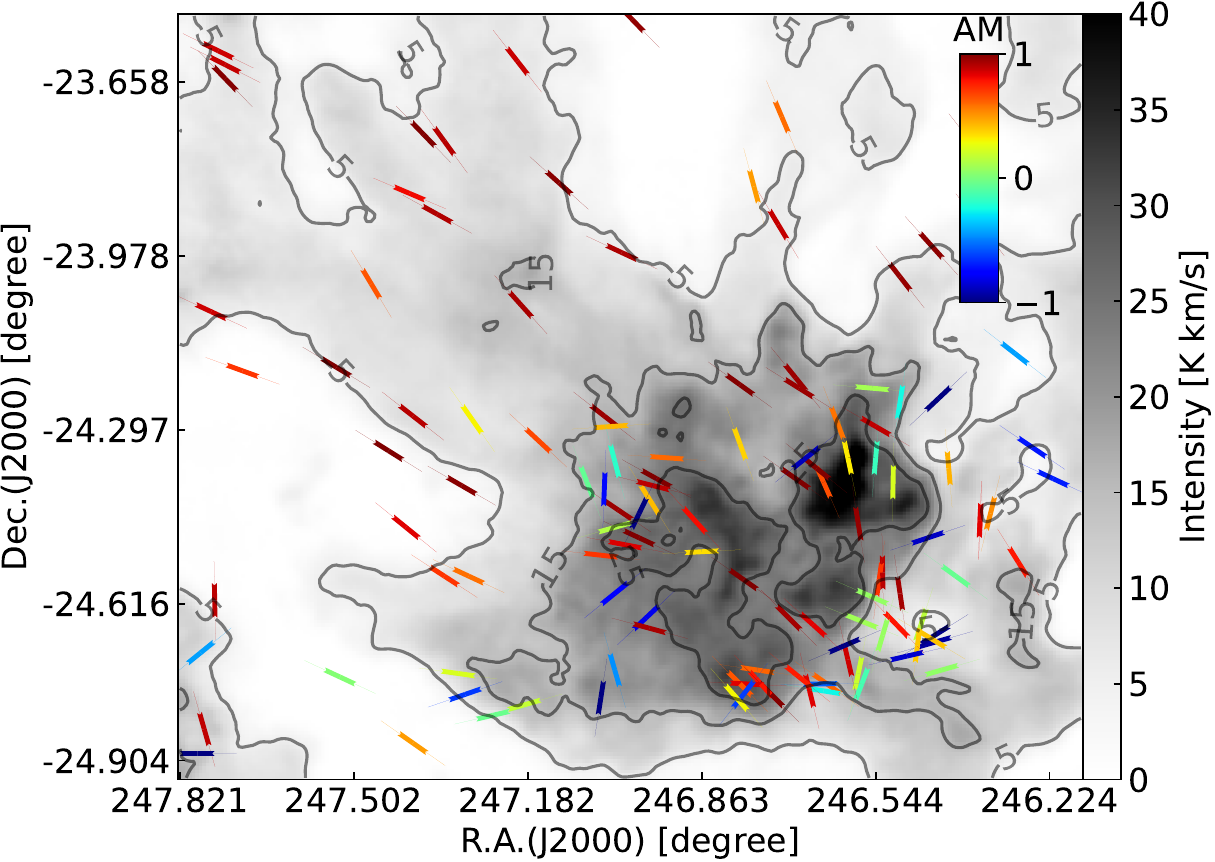}
        \caption{Planck polarization}
        \label{fig:sub2}
    \end{subfigure}
    \caption{Alignment measurement (AM) maps of the magnetic field inferred by stellar polarization compared to the magnetic field orientation inferred by the VGT using $^{12}$CO (top) and the magnetic field inferred by Planck polarization (bottom).}
\label{fig:VGT AM Maps}
\end{figure}

\subsection{VGT-$^{12}$CO: accounting for the contribution from molecular clouds}
As established in our investigations, while VGT-HI is proficient at reconstructing the magnetic field within 3D space—offering a form of magnetic field tomography—it does not encompass the nuanced contributions and variations of the magnetic fields within molecular clouds. This caveat can be mitigated by integrating VGT with molecular emission lines, such as $^{12}$CO.

Fig.~\ref{fig:VGT Star Maps} presents the magnetic fields inferred by stellar polarization within the L1688 region, compared with magnetic field estimations derived from two distinct methodologies: VGT-$^{12}$CO and Planck dust polarization. Collectively, these measurements suggest that the magnetic fields tend to align with the northeastern low-intensity tails of the region. Nonetheless, within the central, high-density core, the magnetic field configurations become markedly complex. Divergences among the three methods are discernible. The stellar polarization vectors are relatively sparse and exhibit a more disordered pattern of magnetic fields. 

Fig.~\ref{fig:VGT AM Maps} presents the AM between the magnetic field orientations deduced from stellar polarization and those inferred from VGT-$^{12}$CO, alongside a comparison with Planck polarization. For VGT-$^{12}$CO, regions of low intensity predominantly show positive AM values, suggesting alignment. In contrast, high-intensity clumps, especially at the cloud's center, exhibit a mix of both positive and negative AM values. However, it is within these denser regions that the most significant perpendicular orientations—or misalignments—of the magnetic fields are observed.

Several factors may contribute to this misalignment. Primarily, stellar polarization may include influences from both the foreground and background, which are not represented in VGT-$^{12}$CO measurements. This is further corroborated by the alignment discrepancies observed between Planck and stellar polarization (see Fig.~\ref{fig:VGT AM Maps}), where Planck data reflect the cumulative magnetic field integrated along the LOS. Additionally, VGT-$^{12}$CO tends to trace magnetic fields within specific volume densities, as optically thick $^{12}$CO generally maps to a critical volume density of 
10$^{2}$~cm$^{-3}$ \citep{2011piim.book.....D}, whereas stellar polarization, driven by dust grain alignment with magnetic fields, might encompass contributions from denser regions \citep{2019MNRAS.482.2697S,2021ApJ...908..218H}. Despite this, the overall agreement between VGT-$^{12}$CO and Planck data (see \citealt{2023MNRAS.524.4431H} and Figs.~\ref{fig:12CO and Planck} and \ref{fig:12CO and Planck AM Map}) attests that VGT-$^{12}$CO is capable of reliably mapping the magnetic fields associated with diffuse molecular gas.

\begin{figure}
    \centering
    \includegraphics[width=1.0\linewidth]{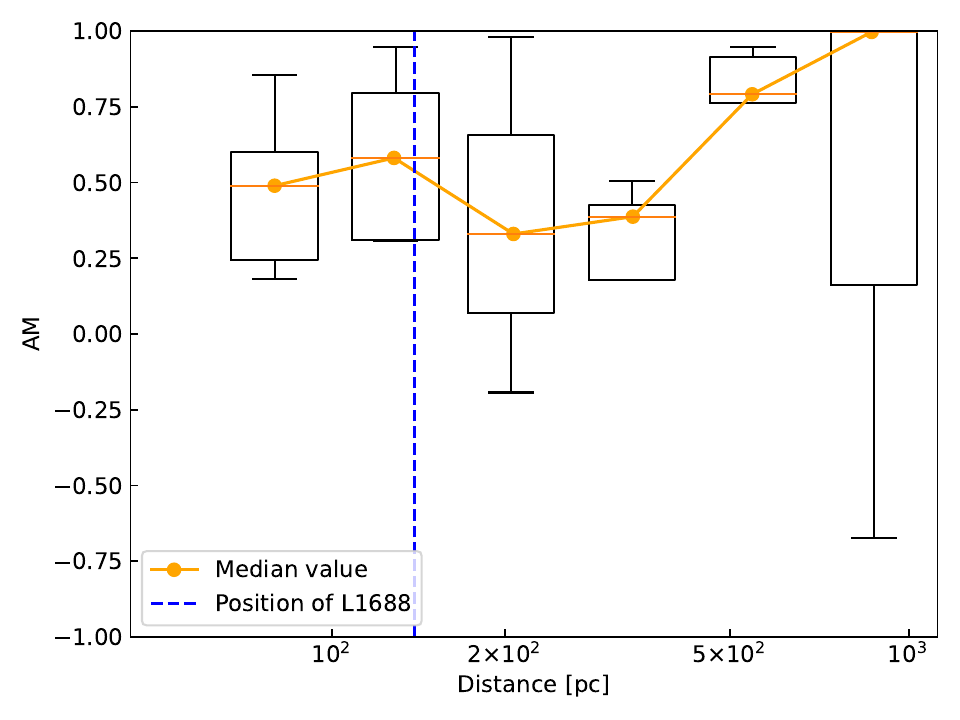}
    \caption{This figure presents the mean AM values of the stellar polarization at certain distances compared to VGT-$^{12}$CO. The upper and lower black lines represent the deviation’s maximum and minimum, respectively. The box gives ranges of the first (lower) and third quartiles (upper) and the orange line represents the median value.}
    \label{fig:AM_distance}
\end{figure}

Moreover, our analysis extends to a comparison between VGT-$^{12}$CO and LOS-decomposed stellar polarization. As depicted in Fig.~\ref{fig:AM_distance}, we observe that the negative AM observed with VGT-HI near L1688 dissipates, and the AM remains consistently positive across all distances. This finding indicates a superior alignment between stellar polarization and VGT-$^{12}$CO within L1688, leading to the conclusion that the negative AM associated with VGT-HI can be attributed to the magnetic field changes within the molecular cloud.

\section{Discussion}
\label{sec:dis}
\subsection{Characterization of POS magnetic fields in 3D}
The quest to map the POS magnetic field in three dimensions within our Galaxy poses a significant scientific challenge. Traditional techniques that trace magnetic fields via dust polarization—such as those employed by the Planck mission—provide invaluable insights into the integrated magnetic fields along the LOS \citep{2020A&A...641A...3P,BG15,10.1111/j.1365-2966.2007.11817.x,2016ApJ...831..159H,2023MNRAS.519.3736H,2023arXiv231017048H}. Despite their utility, these methods alone do not offer the detailed information needed to fully reconstruct the Galactic magnetic field in 3D.

Recent advancements have been made by coupling stellar polarization with precise astrometric data \citep{2012ApJS..200...19C,2019ApJ...872...56P}, and by integrating the VGT-HI with Galactic rotation curves \citep{González-Casanova_2019,2023MNRAS.524.2379H}. These innovative approaches represent a leap forward in our ability to visualize magnetic fields within the three-dimensional expanse of the Galaxy. This study delves into the magnetic fields of the star-forming region L1688, applying the VGT-HI methodology. We use VGT-HI in conjunction with the Galactic rotation curve to create a tomographic map of the magnetic field, which encompasses the regions foreground and background to L1688, and we benchmark the VGT-HI with stellar polarization observations. VGT-HI advantageously provides more extensive field coverage than stellar polarization and we find a significant correlation with stellar polarization data. 

\subsection{Interference from molecular clouds}
A prior investigation into the relationship between VGT-HI and stellar polarization noted a prevalence of negative AM values within regions situated on the Galactic plane and within the spiral arms \citep{González-Casanova_2019}. This pattern was postulated to originate from variations in the magnetic fields associated with molecular clouds. Echoing these findings, our current study also records a decline in the agreement between VGT-HI mapping and stellar polarization near the L1688 region. To unravel the origins of this negative AM, we examined the magnetic field angle variation along the LOS, as shown in Fig.~\ref{fig:Polarization Angle vs. Distance}. Our analysis revealed a change in the stellar polarization angles from a range of [30$^\circ$, 110$^\circ$], with a median at $60^\circ$ east of north, to [0$^\circ$, 180$^\circ$]. This latter range is distinct at L1688's specific LOS position (139 pc). The VGT-HI approach does not account for this rotation, thus making it the probable cause for the observed negative AM.

To rectify this negative AM, we advocate the implementation of VGT-$^{12}$CO, where we noted a statistically significant positive AM that corresponds with both stellar and Planck polarization data (see Fig.~\ref{fig:12CO and Planck AM Map} and Fig.~\ref{fig:AM_distance}). The alignment found in VGT-$^{12}$CO lends credence to the hypothesis that the negative AM detected in VGT-HI stems from the molecular cloud's contribution on stellar polarization. Incorporating VGT-$^{12}$CO, which adeptly maps the local magnetic field within molecular clouds, alongside VGT-HI, offers a promising pathway to synthesize a more comprehensive depiction of the Galactic magnetic field in three dimensions. 

\subsection{Galactic foreground removal in CMB polarization}
Mapping the Galaxy's pervasive 3D magnetic field with greater precision is crucial for effectively modeling the Galactic foreground polarization, which is a well-known confounding factor in discerning B-mode within the CMB polarization \citep{Ade_2023,Macellari_2011,2023arXiv230909978A,Hensley_2018,2020A&A...641A...3P}. Several studies \citep{2019ApJ...887..136C} also proposed the use of HI emission to model the Galactic foreground polarization, while the change of magnetic fields within molecular clouds, as identified in this work, was not accounted for. Here we showed that the combined application of the VGT-HI and VGT-$^{12}$CO presents a promising avenue for accurately modeling the Galactic foreground dust polarization. As shown in Figs.~\ref{fig:HI Velocity Channels}, \ref{fig:AM_distance}, and \ref{fig:12CO and Planck}, this syngengy adeptly delineates the POS magnetic field across the foreground, background, and the L1688 region. Theoretically, this method is scalable and could be applied to broader swathes of the Galaxy, potentially enhancing the fidelity of the Galactic foreground polarization modeling.

\section{Conclusions}
\label{sec:con}
Magnetic fields are an integral component of the ISM, yet mapping these fields, particularly in three-dimensional space, has been a challenging endeavor. In this study, we have employed two distinct methodologies—VGT using HI and $^{12}$CO emission lines, and stellar polarization—to construct a POS magnetic field tomography of the L1688 molecular cloud. Our principal findings are summarized as follows:
\begin{enumerate}
    \item Through the use of stellar polarization, we examined the variation in magnetic field orientation along the LOS. A notable shift in orientation is observed within the vicinity of L1688.
    \item The integrated VGT-HI mapping was found to generally align with the magnetic fields inferred from projected stellar polarization. However, misalignments become evident near L1688 when comparing the integrated VGT-HI with LOS-decomposed stellar polarization.
    \item We observed that integrating VGT-$^{12}$CO mapping with LOS-decomposed stellar polarization mitigates the previously noted misalignment. This, combined with the observed magnetic field variations around L1688, suggested the misalignment between VGT-HI and stellar polarization may be due to the contribution from molecular clouds to stellar polarization.
    \item We have generated a magnetic field tomography of L1688 utilizing VGT-HI in conjunction with the Galactic rotational curve. 
    \item To create a more comprehensive 3D magnetic field map, we propose the integration of VGT-$^{12}$CO to account for the magnetic field within the molecular cloud itself, complementing the magnetic fields obtained from VGT-HI.
\end{enumerate}

\section*{Acknowledgements}
S.T., Y.H., and A.L. acknowledge the support of NASA ATP AAH7546,  NSF grants AST 2307840, and ALMA SOSPADA-016. Financial support for this work was provided by NASA through award 09\_0231 issued by the Universities Space Research Association, Inc. (USRA). This work used SDSC Expanse CPU at SDSC through allocations PHY230032, PHY230033, PHY230091, and PHY230105 from the Advanced Cyberinfrastructure Coordination Ecosystem: Services \& Support (ACCESS) program, which is supported by National Science Foundation grants \#2138259, \#2138286, \#2138307, \#2137603, and \#2138296.

\section*{Data Availability}
The data underlying this article will be shared on reasonable request to the corresponding author.



\bibliographystyle{mnras}
\bibliography{example} 



\appendix
\section{VGT-$^{12}$CO and Planck}
\begin{figure}
    \includegraphics[width=1\linewidth]{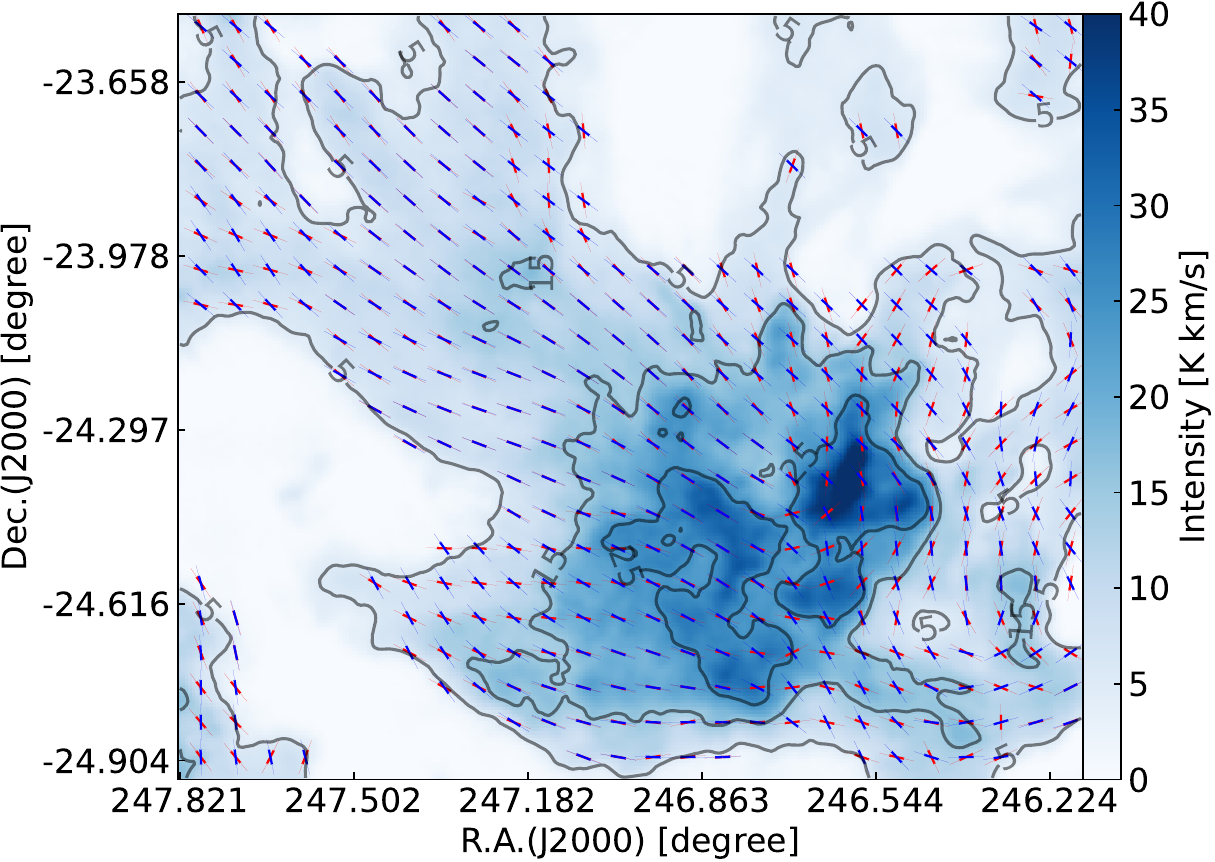}
    \caption{The magnetic field orientation inferred from the VGT using $^{12}$CO with sub block sizes of 20$\times$20 pixels. VGT gradient lines are represented in red and Planck polarization lines are represented in blue. Contours for the map are at 5, 15, and 25 K km/s. Reproduced from \protect\cite{2023MNRAS.524.4431H}.}
    \label{fig:12CO and Planck}
\end{figure}

\begin{figure}
    \includegraphics[width=1\linewidth]{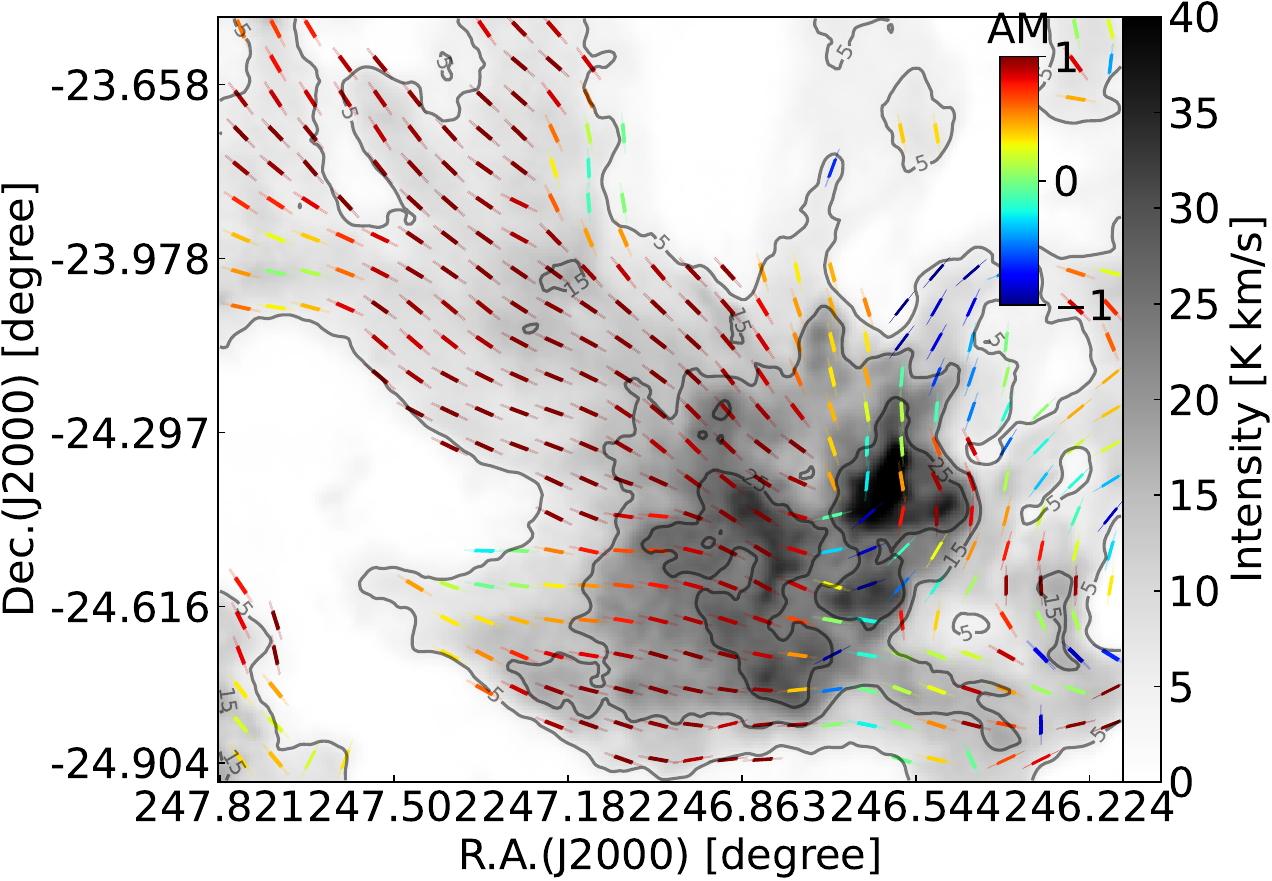}
    \caption{AM map of the magnetic field inferred by VGT using $^{12}$CO compared to the magnetic field orientation inferred by Planck polarization. Reproduced from \protect\cite{2023MNRAS.524.4431H}.}
    \label{fig:12CO and Planck AM Map}
\end{figure}

Figs.~\ref{fig:12CO and Planck} and \ref{fig:12CO and Planck AM Map} present the comparison of the magnetic field inferred from VGT-$^{12}$CO and Planck, as well as their AM map. The maps are reproduced from \cite{2023MNRAS.524.4431H}.


\section{Information about stellar polarization}
Table~\ref{tab:1} consolidates the stellar data utilized in our analysis, encompassing coordinates, polarization angles, and distances. This compilation is drawn from three distinct stellar polarization surveys, as documented in \cite{1976AJ.....81..958V, 1979AJ.....84..199W, 10.1093/mnras/230.2.321}.

\begin{table}
	\centering
	\caption{Information of stars' coordinates (first and second columns), polarization angle (third column), and distance (fourth column) used in this work. The table is
compiled from \protect\cite{1976AJ.....81..958V, 1979AJ.....84..199W, 10.1093/mnras/230.2.321}.}
	\label{tab:1}
	\begin{tabular}{lccr} 
		\hline
		R.A. (J1950) & Dec. (J1950) & $\phi$ [degree] & Distance [pc]\\
		\hline
		16 25 24.3 & -24 27 56.6  & 178 & 136.4\\
		16 26 18.9 & -24 28 19.7 & 9 & 137.5\\
	16 27 46.7 & -24 23 22.1 & 61 & 2733.7\\
        16 27 49.9 & -24 25 40.2 & 32 & 142\\
        16 27 30.2 & -24 27 -43.4 & 42 & NA\\
        16 25 19.2 & -24 26 52.8 & 166 & 140\\
        16 26 10.3 & -24 20 54.8 & 175 & NA\\
        16 26 23.4 & -24 20 59.6 & 12 & 139.2\\
        16 26 45 & -24 23 7.8 & 55 & 110.1\\
        16 27 9.1 & -24 34 8.3 & 55 & 134.6\\
        16 27 27.4 & -24 31 16.4 & 93 & 174\\
        16 26 49 & -24 38 25.2 & 45 & 132.5\\
        16 24 11.4 & -24 59 4 & 53 & 594.1\\
        16 24 8.3 & -24 38 50 & 45 & 110.9\\
        16 24 8.9 & -24 12 30 & 20 & 136.4\\
        16 24 3.8 & -23 44 1 & 15 & 138.4\\
        16 24 52.9 & -24 23 6 & 65 & 114.3\\
        16 25 2.1 & -24 19 54 & 56 & 638.5\\
        16 25 8.9 & -24 9 23 & 52 & 151.8\\
        16 25 46.1 & -23 57 30 & 42 & 400.6\\
        16 25 43.6 & -24 41 21 & 106 & 1103.6\\
        16 25 57.2 & -24 42 35 & 103 & 252.6\\
        16 26 52.9 & -23 55 8 & 31 & 917.6\\
        16 26 36.7 & -24 21 53 & 51 & 650.7\\
        16 26 32.3 & -24 45 53 & 55 & 112.3\\
        16 28 3.1 & -23 58 7 & 65 & 136.1\\
        16 27 57.8 & -23 32 35 & 44 & 424.8\\
        16 28 30 & -23 50 45 & 48 & NA\\
        16 28 18.3 & -24 23 38 & 21 & 144.5\\
        16 28 38.9 & -24 18 52 & 47 & 146.5\\
        16 28 6.1 & -24 35 52 & 129 & NA\\
        16 28 5.2 & -24 44 27 & 16 & 236.4\\
        16 28 10 & -24 56 35 & 68 & 211.5\\
        16 29 5.6 & -24 56 16 & 2 & NA\\
        16 29 11.7 & -24 46 50 & 109 & 81.4\\
        16 29 14.6 & -24 44 34 & 82 & NA\\
        16 29 9.6 & -24 34 0 & 65 & 147.1\\
        16 29 21.1 & -24 33 58 & 57 & 81.1\\
        16 29 7.6 & -24 16 38 & 35 & NA\\
        16 29 34.6 & -24 16 28 & 52 & NA\\
        16 29 23.4 & -23 53 53 & 61 & 746.5\\
        16 29 36.6 & -23 51 37 & 67 & 715.5\\
        16 29 20.6 & -23 46 1 & 36 & 608.2\\
        16 29 30.2 & -23 45 4 & 44 & 530\\
        16 30 1.2 & -24 58 52 & 108 & NA\\
        16 30 1.2 & -24 58 52 & 84 & NA\\
        16 30 6.7 & -24 44 59 & 65 & NA\\
        16 30 7.7 & -24 11 1 & 59 & 376.4\\
        16 31 1.4 & -24 36 43 & 1 & 146.1\\
        16 30 49.5 & -24 11 11 & 70 & 126.6\\
        16 31 0.1 & -23 36 11 & 64 & NA\\
        16 31 25 & -24 7 32 & 57 & 571\\
        16 31 49.5 & -24 17 57 & 44 & 361.1\\
        16 25 24.3 & -24 27 56 & 267 & 244 \\
		\hline
	\end{tabular}
\end{table}

\section{HI spectrum}
\begin{figure*}
    \centering
    \includegraphics[width=0.75\linewidth]{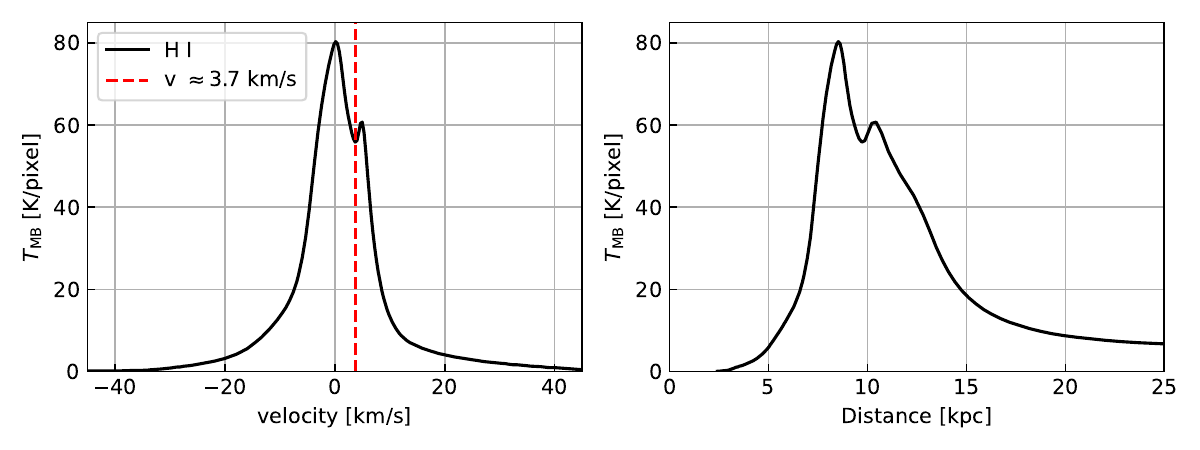}
    \caption{Both figures present the intensity of HI in units of K/pixel. The figure on the left presents the intensity of HI at different velocity channels measured in km/s and the figure on the right presents the intensity of HI at different distances measured in kpc.}
    \label{fig:HI_spectrum}
\end{figure*}

Fig.\ref{fig:HI_spectrum} displays the HI emission spectrum in the direction of the L1688 molecular cloud. We applied the Galactic rotation curve to translate the spectral velocity data into distances from the Galactic center (see \S~\ref{sec:method} for details).

\section{Column Density Maps}
\label{app:D}
\begin{figure*}
    \centering \includegraphics[width=0.55\linewidth]{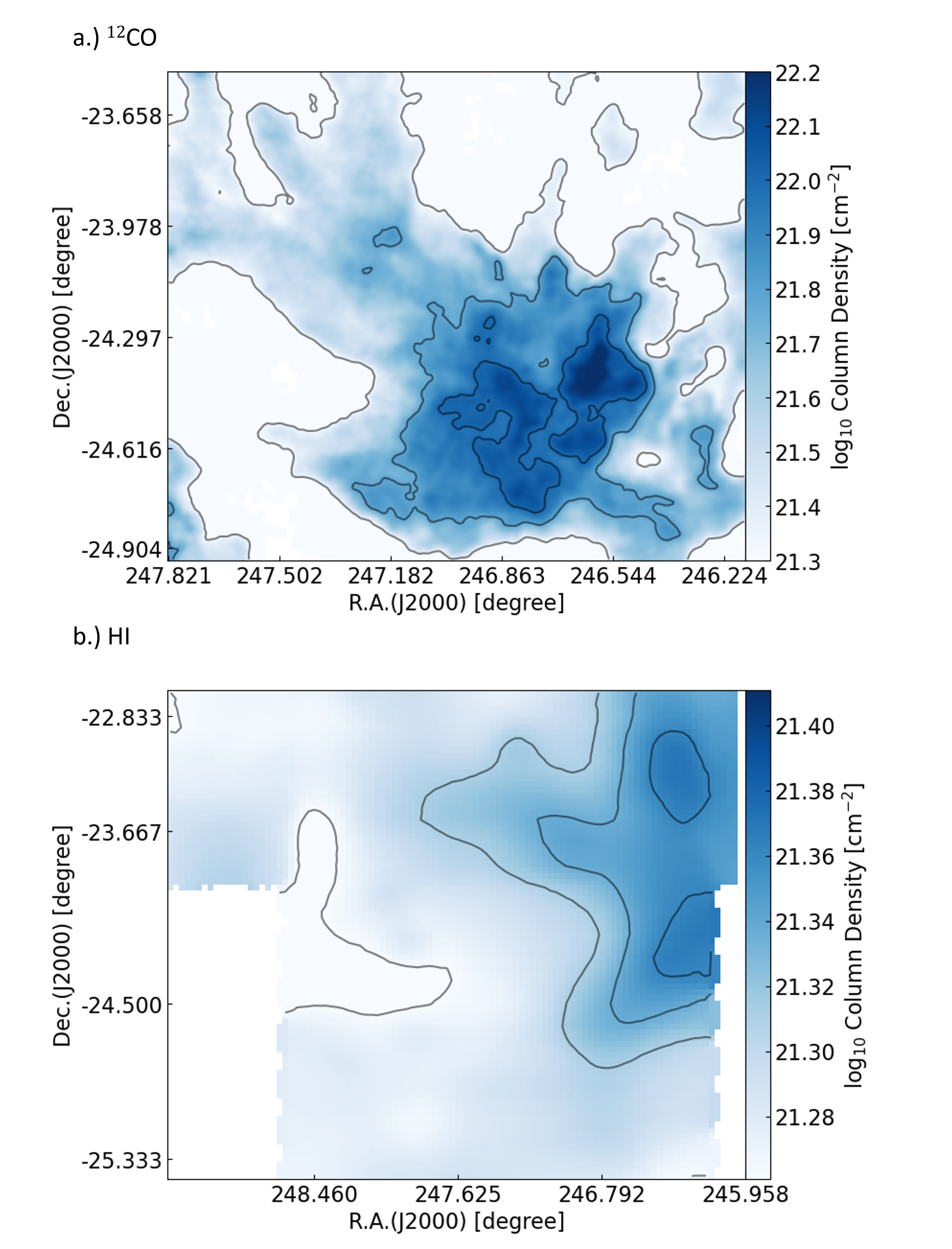}
    \caption{The two figures display the hydrogen column density maps calculated from $^{12}$CO (top) and HI (bottom).}
    \label{fig:Column Density Maps}
\end{figure*}

In Fig.~\ref{fig:Column Density Maps}, we calculated the hydrogen column density, which is well-mixed with grains along the LOS. The column density calculated from H I tells the contribution 
the foreground/background, while the column density calculated from $^{12}$CO represents the column density from the molecular L1688 region. 

\bsp	
\label{lastpage}
\end{document}